\documentclass[english]{llncs}
\usepackage{lmodern}

\usepackage[T1]{fontenc}
\usepackage[latin9]{inputenc}
\usepackage{babel}
\usepackage{array}
\usepackage{verbatim}
\usepackage{varioref}
\usepackage{multirow}
\usepackage{amsmath}
\usepackage{amssymb}
\usepackage[unicode=true,
 bookmarks=false,
 breaklinks=true,pdfborder={0 0 0},backref=false,colorlinks=false]
 {hyperref}
\hypersetup{pdftitle={Why Nominal-Typing Matters in Mainstream OOP},
 pdfauthor={Moez A. AbdelGawad},
 pdfsubject={Nominal-Typing in OOP},
 pdfkeywords={Nominal-Typing, OOP, NOOP, Structural-Typing, Java, C\#, Scala}}

\makeatletter

\newcommand{\noun}[1]{\textsc{#1}}
\providecommand{\tabularnewline}{\\}

\numberwithin{equation}{section}
\numberwithin{figure}{section}
\newcommand{\code}[1]{\texttt{#1}}

\makeatother

\begin{document}

\title{\global\long\def\NOOP{\mathbf{NOOP}}
A Comparison of $\NOOP$ to Structural Domain-Theoretic Models of
OOP}

\author{Moez A. AbdelGawad\\
\code{moez@cs.rice.edu}}

\institute{College of Mathematics and Econometrics, Hunan University\\
Changsha 410082, Hunan, P.R. China\\
Informatics Research Institute, SRTA-City\\
New Borg ElArab, Alexandria, Egypt\\
}
\maketitle
\begin{abstract}
Mainstream object-oriented programming languages such as Java, C\#,
C++ and Scala are all almost entirely nominally-typed. $\NOOP$ is
a recently developed domain-theoretic model of OOP that was designed
to include full nominal information found in nominally-typed OOP.
This paper compares $\NOOP$ to the most widely known domain-theoretic
models of OOP, namely, the models developed by Cardelli and Cook%
, which were structurally-typed models. Leveraging the development
of $\NOOP$, the comparison presented in this paper provides a clear
and precise mathematical account for the relation between nominal
and structural OO type systems.
\end{abstract}

\section{Introduction}

\global\long\def\SOOP{\mathbf{SOOP}}
\global\long\def\dom#1{\mathcal{#1}}
\global\long\def\strfunarr{\multimap\!\rightarrow}
\global\long\def\Sig#1{\mathsf{s_{#1}}}
\global\long\def\subsign{\trianglelefteq}
\global\long\def\rec{\multimap}
\global\long\def\ext{\blacktriangleleft}
The first mathematical models of object-oriented programming (OOP)
to gain wide-spread recognition were structural models. Being structural,
objects were viewed in these models as being mere records. Object
types, in accordance, were viewed as record types, where the type
of an object specifies the structure of the object, meaning that object
types carry information on the names of the members of objects\emph{
}(\emph{i.e.}, fields and methods), and, inductively, on the (structural)
types of these members. The model of OOP developed by Cardelli in
the eighties of last century, and later enhanced by Cook and others,
is an example of a structural model of OOP. Examples of structurally-typed
OO languages include lesser-known languages such as O'Caml~\cite{OCamlWebsite},
Modula-3~\cite{Cardelli89modula}, Moby~\cite{Fisher1999}, PolyTOIL~\cite{Bruce2003},
and Strongtalk~\cite{Bracha1993}.

Despite the popularity of the structural view of OOP among programming
languages researchers, many industrial-strength mainstream OO programming
languages are nominally-typed. Examples of nominally-typed OO languages
include well-known languages such as Java~\cite{JLS14}, \noun{C\#~\cite{CSharp2015},}
\noun{C++}~\cite{CPP2011}, and Scala~\cite{Odersky14}. In nominally-typed
OO languages, objects and their types are nominal, meaning that objects
and their types carry\emph{ class names information} (also called
nominal information) as part of the meaning of objects and of their
types, respectively.

In pure structurally-typed OO languages, nominal information is not
used as part of the identity of objects and of their types during
static type checking%
{} nor is nominal information available at runtime%
\footnote{Given that most industrial-strength OO languages are statically-typed,
in this work we focus on nominal and structural \emph{statically}-typed
OO languages. A discussion of statically-typed versus dynamically-typed
OO languages (including the non-well-defined so-called ``duck-typing''),
and the merits and demerits of each, is beyond the scope of this work.
The interested reader should check~\cite{Meijer2004}.}. Accordingly, nominal information is missing in all structurally-typed
models of OOP.

OOP was in its early days at the time the first mathematical models
of OOP were developed, in the eighties of last century. Functional
programming was the dominant programming paradigm among programming
languages researchers at that time---and largely still is today. As
such, the role of nominality of objects and of their types (\emph{i.e.},
the inclusion of nominal information in their identities) in the semantics
of mainstream OOP was not widely appreciated, and nominal OO type
systems remain under-researched. $\NOOP$~\cite{NOOP,NOOPsumm} is
a recently developed domain-theoretic model of OOP that addresses
this shortcoming. To the best of our knowledge, $\NOOP$ is so far
the only domain-theoretic model of OOP to include \emph{full} class
names information as found in mainstream nominally-typed OO programming
languages. %
In this paper, we compare $\NOOP$ to other well-known structural
domain-theoretic models of OOP.

This paper is structured as follows. First, we discuss related work---including
the history of modeling OOP---in Section~\ref{sec:Related-Work}.
As an appetizer for the following comparison, a fundamental technical
difference between pure nominally-typed OO languages and pure structurally-typed
OO languages is discussed in Section~\ref{sub:Type-Names,Circularity,Bin-Methods}.
In Section~\ref{sec:Nominally-Typed-Denotational-Mod} we then compare
the nominal mathematical view of OOP to the structural mathematical
view of OOP%
{} by presenting a comparison between $\NOOP$ and the structural models
of OOP constructed by Cardelli and enhanced by Cook and others. We
conclude in Section~\ref{sec:Future-Work} by summarizing our findings,
making some final remarks, and discussing some possible future research.

\section{\label{sec:Related-Work}Related Work}

Even though object-oriented programming emerged in the 1960s, and
got mature and well-established in mainstream software development
in the late 1980s, the differences between nominally-typed and structurally-typed
OO programming languages started getting discussed by programming
languages (PL) researchers only in the 1990s~\cite{Magnusson91,Porter92,Thorup99}.
In spite of these early research efforts, the value of nominal typing
and nominal subtyping to mainstream OO developers did not get the
full attention of the PL research community until around the turn
of the century.

In the eighties, while OOP was in its infancy, Cardelli built the
first denotational model of OOP~\cite{Cardelli84,Cardelli88}. Cardelli's
work was pioneering, and naturally, given the research on modeling
functional programming extant at that time, the model Cardelli constructed
was a structural denotational model of OOP.\footnote{Quite significantly, Cardelli in fact also hinted at looking for investigating
nominal typing~\cite[p.2]{Cardelli87}. Sadly, Cardelli's hint went
largely ignored for years, and structural typing was rather \emph{assumed}
superior to nominal typing instead, particularly after the publication
of Cook et al.'s and Bruce et al.'s work.} In the late 1980s/early 1990s, Cook and his colleagues worked to
improve on Cardelli's model, leading them to break the identification
of the notions of inheritance and subtyping~\cite{CookDenotational89,Cook1989,CookInheritance90}%
. Unlike Cardelli, Cook emphasized in his work---as we discuss in
more detail in Section~\ref{sub:-Compared-toCOOK}---the importance
of \emph{self-references} in OOP, at the value level (\emph{i.e.},
self variables, such as \code{this}) and at the type level (\emph{i.e.},
self-type variables). 

In 1994, Bruce et al. presented a discussion of the problem of binary
methods in OOP~\cite{BruceBinary94}. Later, Bruce, and Simons, also
promoted the structural view of OOP in a number of publications (\emph{e.g.},
\cite{BruceFoundations02} and~\cite{SimonsTheory02}) and they promoted
conclusions based on this view. However, the deep disagreement between
these conclusions (such as breaking the correspondence between inheritance
and subtyping) and the fundamental intuitions of a significant portion
of mainstream OO developers persisted~\cite{InhSubtyNWPT13,AbdelGawad2015}.

Under the pressure of this disagreement, some PL researchers then
started in the late 1990s/early 2000s stressing the significance of
the differences between nominally-typed OOP and structurally-typed
OOP, and they started acknowledging the practical value of nominal
typing and nominal subtyping (see~\cite{AbdelGawad2015,OOPOverview13}
for more details) and asserted the need for more research on studying
nominal OO type systems~\cite{TAPL}. Accordingly, some attempts
were made to develop OO languages that are both nominally- and structurally-typed~\cite{Findler04,Ostermann08,Gil08,Malayeri08,Malayeri2009,Odersky14}.\footnote{Multiple dispatch (see~\cite{Chambers92,Boyland97,Clifton06}), also,
was discussed (\emph{e.g.}, in~\cite{BruceBinary94}) as a possible
solution to the problem of binary methods.} However, at least in the eyes of mainstream OO developers, these
hybrid languages have far more complex type systems than those of
OO languages that are either purely nominally-typed or purely structurally-typed
(see discussion in Section~\ref{sub:NOOPvsCardelli}).

As to \emph{operational} mathematical models of OOP, Abadi and Cardelli
were the first to present such a model~\cite{SemObjTypes94,TheoryOfObjects95}.
Their model also had a structural view of OOP. However, operational
models of nominally-typed OOP got later developed. In their seminal
work, Igarashi, Pierce, and Wadler presented Featherweight Java (FJ)~\cite{FJ/FGJ}
as an operational model of a nominally-typed OO language. Even though
FJ is not the first operational model of nominally-typed OOP (for
example, see~\cite{drossopoulou99},~\cite{nipkow98} and~\cite{flatt98,flatt99}),
yet FJ is the most widely known operational model of (a tiny core
subset of) a nominally-typed mainstream OO language, namely Java.
The development of FJ and other operational models of nominally-typed
OOP marked a strong focus on studying nominal-typing in OO languages,
thereby departing from earlier disregard of it.

These developments later motivated the construction of $\NOOP$.
Featherweight Java (FJ) in fact offers the closest research to $\NOOP$
since it offers a very clear operational semantics for a tiny nominally-typed
OO language. It is worth mentioning that $\NOOP$, as a more foundational
domain-theoretic model of nominally-typed OO languages (\emph{i.e.},
that has fewer assumptions than FJ), provides a denotational\emph{
}justification for the inclusion of nominal information in FJ. The
inclusion of nominal information in $\NOOP$ is crucial for proving\emph{
}the identification of inheritance and subtyping in nominally-typed
OOP. In\noun{ FJ~\cite{FJ/FGJ}}, rather than being proven as a consequence
of nominality, the identification of inheritance and subtyping was
taken as an\emph{ }assumption. $\NOOP$ also allows discussing issues
of OOP such as type names, `self-types' and binary methods on a
more foundational level than provided by operational models of OOP.
The more abstract description of denotational models results in a
conceptually clearer understanding of the programming notions described,
as well as of the relations between them.\footnote{It is worthy to also mention that $\NOOP$ was developed, partially,
in response to the technical challenge Pierce (an author of FJ) presented
in his LICS'03 lecture~\cite{Pierce03} in which Pierce looked for
precising the relation between structural and nominal OO type systems
(notably,\emph{ after} the development of FJ was concluded).}

Finally, related to our work is also the dissatisfaction some researchers
expressed about possible misunderstandings extant in the PL research
community, and about the (mal)practices based on these misunderstandings
when PL researchers study object-oriented programming languages in
particular. Given the different basis for deriving data structuring
in functional programming (based on standard branches of mathematics)
and in object-oriented programming (based on biology and taxonomy)~\cite{Cardelli84,Cardelli88},
some PL researchers have expressed dissatisfaction with assuming that
the views of programming based on researching functional programming
(including a view that assumes structural typing) may apply without
qualifications to object-oriented programming. In addition to pointing
out the importance of distinguishing between nominal typing and structural
typing, MacQueen~\cite{MacQueenMLOO02}, for example, has noted many
mismatches between Standard ML~\cite{Milner97} (a popular functional
programming language) and class-based OO languages such as Java and
C++. Later, Cook~\cite{cook-revisited} also pointed out differences
between objects of OOP and abstract data types (ADTs), which are commonly
used in functional programming.\footnote{We consider these research results as running in a similar vein as
ours, since they somewhat also point to some mismatches between the
theory and practice of programming languages---theory being mathematics-based,
functional, and structurally-typed, and practice being biology/taxonomy-based,
object-oriented, and nominally-typed.}\textsuperscript{,}\footnote{Yet another research that is also somewhat similar to the one we present
here, but that had different research interests and goals, is that
of Reus and Streicher~\cite{Reus02,Reus02a,Reus03}. In~\cite{Reus03},
an untyped denotational model of class-based OOP is developed. Type
information is largely ignored in Reus and Streicher's work (in particular,
members of objects have no type signatures) and some minimal amount
of nominal information is included with objects only to support analyzing
OO dynamic dispatch. This model was developed to analyze mutation
and imperative features of OO languages and for developing specifications
of OO software and the verification of its properties~\cite{Reus03}.
Analyzing the differences between structurally-typed and nominally-typed
OO type systems was \emph{not} a goal of Reus and Streicher's research.
Despite the similarity of $\NOOP$ and the model of Reus and Streicher,
we thus make no further mention of Reus and Streicher's model in
this paper due to its different interests and goals, and due to the
fundamentally different nature of $\NOOP$ compared to their model
(\emph{i.e.}, $\NOOP$ including all essential class names information
inside objects versus Reus and Streicher's model lacking most of
this information.)}

\section{\label{sub:Type-Names,Circularity,Bin-Methods}Type Names, Type Contracts,
Recursive Types and Binary Methods}

From the point of view of OO developers and OO language designers,
there are many technical differences between nominally-typed OO languages
and structurally-typed OO languages. We discuss these in brief in
this section. (A more detailed discussion is presented in~\cite{AbdelGawad2015}
and \cite{OOPOverview13}.)

\medskip{}

\noindent \textbf{\emph{Type Names and Behavioral Type Contracts}}
A fundamental technical difference between nominally-typed OO type
systems and structurally-typed OO type systems is how the two approaches
view type names. In structurally-typed OO languages, type names
are viewed as being names for type variables that \emph{abbreviate
type expressions} (\emph{i.e.}, are ``shortcuts''). As such, the
use of type names in structurally-typed OO languages is not always
necessary, but type names are useful as abbreviations and they are
even necessary for defining recursive type expressions. As variable
names, however, recursive type names in structurally-typed OO languages
(such as the name of a class when used inside the definition of the
class---which gets interpreted as ``self-type'') get \emph{rebound}
to different types upon type inheritance, and they get rebound to
types that, if they were subtypes, could break the contravariant subtyping
rule of method parameter types (and, thus, break the type safety of
structurally-typed OO languages). Structurally-typed OO languages
resolve this situation by breaking the correspondence between type
inheritance and subtyping.

In nominally-typed OO languages, on the other hand, the nominality
of types means type names are viewed as part of the identity and meaning
of type expressions, since type names in these languages are associated
with public formal or informal \emph{behavioral} \emph{contracts}.\footnote{In well-designed OO programs, each class (and interface and trait,
in languages that support these notions) has associated contracts
describing the behavior of objects of the class (its instances). The
contracts include an invariant (a predicate) for the values of class
fields, and a contract for each method stipulating what conditions
the inputs should satisfy and what output condition should hold over
the value returned by the method, as well as side effects that have
been performed on this and perhaps other objects passed as arguments
to the method. (The output predicate may mention the values of arguments
and in such case is often called an input-output predicate.) In practice,
class contracts are typically informal, may be incomplete, and are
usually expressed only in code documentation. (See~\cite{AbdelGawad2015}
for a longer, detailed and deeper discussion of the association of
type names with contracts, and of the import of this association to
mainstream OO developers.)} Being names for public, and thus \emph{fixed}, contracts means that,
in nominally-typed OO languages, type names cannot be treated as variable
names. In nominally-typed OO languages, thus, type names have fixed
meanings that do \emph{not} change upon inheritance. Further, in
these languages the fixed type a type name is bound to does \emph{not}
break the contravariant subtyping of method parameters when the method
and its type get inherited by subtypes (types corresponding to subclasses/sub\-interfaces).
As such, in nominally-typed OOP it is not necessary to break the identification
of type inheritance with subtyping.\medskip{}

\noindent \textbf{\emph{Recursive Types}} Further, in class-based
OOP, a class (or interface or trait, in languages that support these
notions) can directly refer to itself (using class/inter\-face/trait
names) in the signature of a field, or the signature of a method parameter
or return value, where the class name is used also as a type name.
This kind of reference is called a type \emph{self-reference}, \emph{recursive
reference}, or, sometimes, \emph{circular} \emph{reference}. Also,
mutually-dependent classes, where a class refers to itself indirectly
(\emph{i.e.}, via other classes), are allowed in class-based OOP.
As Pierce noted~\cite{TAPL}, nominally-typed OO languages allow
readily expression of mutually-dependent class definitions. Since
objects are characterized as being self-referential values (according
to Cook~\cite{cook-revisited}, objects are `autognostic'), and
since self-referential values can be typed using recursive types~\cite{MPS},
there is wide need for recursive type definitions in mainstream OOP.
As such, direct and indirect circular type references are quite common
in mainstream OOP~\cite{cook-revisited}. The ease by which recursive
typing can be expressed in nominally-typed OO languages is one of
the main advantages of nominally-typed OOP.\footnote{According to Pierce~\cite[p.253]{TAPL}, ``The fact that recursive
types come essentially for free in nominal systems is a decided benefit
{[}of nominally-typed OO languages{]}.''}  

In the comparison of nominal and structural mathematical models of
OOP in Section~\ref{sec:Nominally-Typed-Denotational-Mod} we will
see that, in accordance with their different views of type names,
self-referential class references are viewed differently by nominally-typed
models of OOP than by structurally-typed models of OOP. The different
views of circular class references are behind nominal models of OOP
leading to a different conclusion about the relation between inheritance
and subtyping than the conclusion reached based on structural models.\medskip{}

\noindent \textbf{\emph{Binary Methods}} From the point of view of
OO developers, the difference between the nominal and the structural
views of type names in OOP demonstrates itself, most prominently,
in the different support and the different treatment provided by OO
languages to what are usually called ``binary methods''. In OOP,
a `binary method' is defined as a method that takes a parameter
(or more) of the same type as the class the method is declared in~\cite{BruceBinary94}.
``The problem of binary methods'' and requiring them to be supported
in OO languages was a main motivation behind structural models of
OOP leading to inheritance and subtyping not being identified (\emph{i.e.},
as not being in a one-to-one correspondence)~\cite{CookInheritance90}.
As explained above, given their view of type names as type variable
names, structurally-typed OO languages require the self-type of the
argument of a method---where the method is identified as a binary
method, and upon inheritance of the method by a subclass of the class
the method is first declared in---to be that of the type corresponding
to the \emph{subclass}.

Nominally-typed OO languages, on the other hand, with their fixed
interpretation of type names, treat a method taking in an argument
of the same class as that in which the method is declared like any
other method, \emph{i.e.}, needing no special treatment. As such,
nominally-typed OO languages guarantee that the type of the input
parameter of a method that approximates a binary method is a \emph{supertype}
of its type if it were a true binary method.

Nominally-typed OO languages, thus, offer a somewhat middle-ground
solution between totally avoiding binary methods and overly embracing
them (as pure structurally-typed OO languages do). Given that the
meaning of types names in nominally-typed OO languages does not change
upon inheritance, these languages provide methods whose type, upon
inheritance, only approximates the type of true binary methods. Nominally-typed
OO languages do not quite support binary methods, but, for good reasons
(\emph{i.e.}, so as to not break the identification of inheritance
of contracts and subtyping, nor lose other advantages of nominal typing~\cite{AbdelGawad2015}),
offer only a good approximation to binary methods. Given that the
type of the parameter does not change in subclasses, the degree of
approximation (if the method was indeed a true binary method) gets
lesser the deeper in the inheritance hierarchy the method gets inherited.\footnote{With the introduction of generics~\cite{JLS14,CSharp2015,Odersky14,Bank96,Agesen97,Bracha98,FJ/FGJ},
and `F-bounded generics' (the nominal counterpart of F-bounded polymorphism~\cite{CanningFbounded89,Baldan1999,Greenman2014})
in particular, nominally-typed OO languages provided better support
for \emph{true} binary methods while keeping the identification of
type inheritance with subtyping and other benefits of nominal typing.
It should be noted that the lesser-recognized problem of `spurious
binary methods' in structurally-typed OOP (see~\cite[Section 3.3.1]{AbdelGawad2015})
provides further justification for nominally-typed OO languages being
cautious about fully embracing binary methods by treating a method
that ``looks like'' a binary method as indeed being one. In light
of the spurious binary methods problem, and precluding the use of
F-bounded generics, in our opinion a better approach towards supporting
true binary methods in mainstream OO languages might be by allowing
developers to \emph{explicitly} mark or flag true binary methods as
being such, or, even more precisely, to allow developers to mark specific
arguments of methods as being arguments that `need to be treated
as those of true binary methods.'}

\section{\label{sec:Nominally-Typed-Denotational-Mod}Nominally-Typed versus
Structurally-Typed Models of OOP}

To see how nominality and nominal typing affects mathematical views
of OOP, we compare $\NOOP$, as a nominally-typed denotational model
of OOP, to the most widely known structural model of OOP---the one
constructed and presented by Cardelli~\cite{Cardelli84,Cardelli88},
and extended, analyzed and promoted by others such as Cook~\cite{CookDenotational89,Cook1989,CookInheritance90},
Bruce~\cite{BruceFoundations02} and Simons~\cite{SimonsTheory02}.

Even though unnamed by their authors, for ease of reference in this
paper we call Cardelli's model $\SOOP$, for Structural OOP, while
calling the extension of $\SOOP$ by Cook et al. $\mu\SOOP$ (due
to its inclusion of recursive types). As we discussed earlier, $\NOOP$
is the first domain-theoretic model of OOP to include full nominal
type information found in nominally-typed OOP. The construction of
$\NOOP$ is presented in~\cite{NOOP}, and is summarized in~\cite{NOOPsumm}.
In the following sections we first compare $\NOOP$ to $\SOOP$ then
compare it to $\mu\SOOP$.

\subsection{\label{sub:NOOPvsCardelli}$\protect\NOOP$ Compared to $\protect\SOOP$}

The model of OOP developed by Cardelli in the 1980s~\cite{Cardelli84,Cardelli88}
was the first denotational model of OOP to gain widespread recognition.
In his pioneering and seminal work Cardelli, according to him himself,
had a goal of `unifying functional programming and object-oriented
programming'~\cite[p.2]{Cardelli88}. %
A domain equation that describes the main features of $\SOOP$ (distilled
to exclude variants. See~\cite[pp.15, 16]{Cardelli88} for the actual
domain equations used by Cardelli)  is
\[
\dom V=\dom B+(\dom V\rightarrow\dom V)+(\dom L\rightarrow\dom V)
\]
where $\dom V$ is the main domain of values, $\dom B$ is a domain
of base values, $\dom L$ is the flat domain of labels, $\rightarrow$
is the standard continuous functions domain constructor, and $+$
is the disjoint summation domain constructor. The distilled $\SOOP$
domain equation expresses the view that values are either base values,
unary functions over values, or \emph{records} (``objects'') modeled
as (infinite) functions from labels to values.

The domain equation describing $\NOOP$ is
\[
\dom O=\mathcal{S}\times(\mathcal{L}\multimap\mathcal{O})\times(\mathcal{L}\multimap(\mathcal{O}^{*}\strfunarr\mathcal{O}))
\]
where the main domain defined by the equation, namely domain $\dom O$,
is the domain of (raw) objects, $\times$ is the strict product domain
constructor, and $\rec$ is the records domain constructor (See~\cite{NOOPsumm}
or~\cite[Chapter 6]{NOOP} for more details on the $\NOOP$ domain
equation). The $\NOOP$ domain equation expresses the view that every
object is a triple of: (1) a class signature closure (\emph{i.e.},
a member of domain $\mathcal{S}$), (2) a fields record (\emph{i.e.},
a member of $\mathcal{L}\multimap\dom O$), and (3) a methods record
(\emph{i.e.},\emph{ }a member of $\mathcal{L}\multimap(\dom O^{*}\strfunarr\dom O),$
where $\strfunarr$ is the strict continuous functions domain constructor,
and $^{*}$ is the finite-sequences domain constructor).

Class signatures and other related constructs are syntactic constructs
that capture \emph{all} nominal (\emph{i.e.}, class/interface/trait
names) information found in objects of mainstream nominally-typed
OO software~\cite{NOOPsumm,NOOP}. Class signatures formalize the
informal notion of `object interfaces'~\cite{AbdelGawad2015,OOPOverview13,NOOP}.
Embedding class signature constructs in objects of $\NOOP$ makes
them nominal objects. It should be noted that consistency conditions
for signature constructs in $\NOOP$~\cite[Section 4]{NOOPsumm}~\cite[Section 5.1]{NOOP}
do not preclude a signature from directly or indirectly referring
to itself in the signature of a field or of a method parameter or
method return value, so as to allow for self-referential types (see
Section~\ref{sub:Type-Names,Circularity,Bin-Methods}.)

A comparison of $\NOOP$ to $\SOOP$ reveals the following fundamental
difference between the two models:
\begin{itemize}
\item $\SOOP$ is a structural model of OOP, that, as explained by its domain
equation, does \emph{not} include nominal information into its objects.
As such, $\SOOP$ views objects as being \emph{essentially records}
(of functions)~\cite[p.3]{Cardelli88}. Due to the lack of nominal
information, the definitions of types of objects and of subtyping,
based on $\SOOP$, are also structural definitions, \emph{i.e.}, ones
that can only respect object structures but that \emph{cannot }respect
the behavioral contracts maintained by objects that are associated
with their type names.
\item $\NOOP$ is a nominal model of OOP, that, via the $\dom S$ component
(for signatures) of its domain equation, \emph{includes full nominal
information} into its objects . As such, $\NOOP$ views objects as
records (of fields and methods) \emph{accompanied by nominal information}
referencing the behavioral contracts maintained by the fields and
methods of the objects. The definition of types of objects and of
subtyping, based on $\NOOP$, can thus be nominal ones, \emph{i.e.,
}ones which \emph{can} respect behavioral contracts associated with
type names in addition to respecting object structures.
\end{itemize}
In the comparison of $\NOOP$ to $\SOOP$ it should also be noted
that the `Inheritance $\Leftrightarrow$ Subtyping' (`inheritance
is subtyping') theorem of $\NOOP$ (\cite[Section�5.3]{NOOPsumm}),
stating the identification of type inheritance with subtyping in nominally-typed
OOP, is very similar to Cardelli's `Semantic Subtyping' theorem
(\cite[Section�11]{Cardelli88}). Cardelli did not model recursive
types, and thus did not handle recursive type expressions (which are
the structural counterpart of self-referential class signatures).
As such, despite the model of Cardelli being a structural model of
OOP, the omission of recursive types enabled Cardelli to easily identify
an inaccurate ``structural'' notion of inheritance with a structural
definition of subtyping and prove their one-to-one correspondence\footnote{In his work, Cardelli, informally and somewhat implicitly, defined
inheritance as structural subtyping between (record) type expressions.
Demonstrating the strong influence of functional programming on Cardelli's
model, Cardelli even argued for expanding the definition of inheritance
to include some notion of ``inheritance'' between function types
(by which it seems Cardelli really meant subtyping, since Cardelli
did not suggest any code sharing).}.

Other tangential differences that are noted in the comparison between
$\NOOP$ and $\SOOP$ include:
\begin{enumerate}
\item $\SOOP$ models records as infinite functions, with only an informal
restriction on the functions that requires the functions to map a
cofinite set of input labels---\emph{i.e.}, all but a finite number
of labels---to the value \code{wrong}. $\NOOP$, on the other hand,
models the record component of objects using the $\rec$ (`rec')
domain constructor which constructs records as tagged \emph{finite}
functions. Domain constructor $\rec$, even though having similarity
to some other earlier-developed domain constructors, was particularly
developed to let $\NOOP$ model mainstream OOP more accurately. Because
of using $\rec$, the $\NOOP$ domain of objects formally includes
no ``junk'' (\emph{i.e.}, unnecessary) infinite records as those
found in the formal definition of $\SOOP$.
\item \label{enu:pureOO}Given its attempt to unify FP and OOP, $\SOOP$
allows functions as first-class values in its main domain of values.
As such, $\SOOP$ is not a pure-OO model of OOP. $\NOOP$, on the
other hand, is a pure-OO model of OOP. Every value in the main domain
of $\NOOP$ is an object. To model methods and records, $\NOOP$ uses
functional domains, but they are used only as auxiliary domains.
\item Functions (used to model methods) in $\SOOP$ are unary functions
that take exactly one argument---an element of domain $\dom V$. $\SOOP$
thus requires `currying' to model multi-ary functions and methods.
In $\NOOP$, on the other hand, sequences\emph{ }of objects are used
as method arguments to model multi-ary methods more precisely (\emph{i.e.},
without the need for currying, which is not commonly familiar to mainstream
OOP developers as it is to FP developers, and thus, inline with the
previous point, also without the need for functions/methods to be
first-class values).
\item $\SOOP$ uses the same namespace for fields and methods of records,
disallowing a field in a record to have the same name as a method
in the record. $\NOOP$, on the other hand, aims to mimic mainstream
OO languages more closely, and thus it uses two records as separate
components inside objects to give fields and methods separate namespaces.
A field and a method in a $\NOOP$ object can thus have the same name
without conflict (method overloading, however, where two methods inside
an object can have the same name, is supported neither by $\SOOP$
nor by $\NOOP$).\footnote{To put research on structural OOP on a more rigorous footing, and
as a step towards the construction of $\NOOP$, we constructed $\mathbf{COOP}$---\cite[Ch. 4]{NOOP}
and~\cite[Sec. 4]{DomThSummCOOP14}---as a simple structural domain-theoretic
model of OOP that dealt with the first three of the four tangential
differences between $\NOOP$ and $\SOOP$.}
\end{enumerate}

\paragraph{\textbf{Nominal vs. Structural vs. Hybrid Typed OO Languages}}

It is worthy to mention here that the fundamental `structural versus
nominal' difference between $\SOOP$ and $\NOOP$ has profound implications
on comparing nominally-typed OO languages to structurally-typed OO
languages, and to hybrid OO languages that try or claim to support
both nominal and structural typing.

First, it is clear that supporting nominal typing in an OO language
with a structural view of objects is impossible, since the nominal
information stripped by the structural view of objects is irrecoverable
from the structure of the objects. Second, due to the association
of type names to behavioral contracts, it is clear nominal typing
is closer to semantic/behavioral typing than structural typing is
(More discussion of contracts and semantic typing is presented in~\cite{AbdelGawad2015}).

Thirdly, from the definition of $\NOOP$ it is clear also that, if
needed, it is easy to define structural types on a domain of nominal
objects. The definition of these types can be done in $\NOOP$ by
ignoring nominal information, as is done in ``hybrid'' OO languages
such as Scala, SmallTalk, Whiteoak and Unity. The definition of these
structural types in this case is not the same as for an OO language
based on a structural view of objects and modeled by $\SOOP$, since
objects of the defined structural types will still carry nominal information
at run-time (ready to be used during software run-time, such as in
type casting operations and \code{instanceof} type tests). Structural
OO languages that support a structural view of objects are fundamentally
different than nominal languages because objects in such languages,
as modeled by $\SOOP$, are plain records (and thus without any reference
to behavioral class contracts), which is \emph{not} true in OO languages
that try to support both nominal and structural types.\footnote{A further reason we do not believe hybrid languages, such as Scala~\cite{Odersky14},
SmallTalk~\cite{Smalltalk98}, Whiteoak~\cite{Gil08} and Unity~\cite{Malayeri08},
indeed provide true or full support for structural typing is that
these languages do not quite support \emph{recursive }structural types
(varying between having reluctant/weak support to having no support
for them at all). As discussed in Section~\ref{sub:Type-Names,Circularity,Bin-Methods},
recursive types are essential for serious OO programming. As demonstrated
by Cook's work (which we discuss in more detail in the next section),
supporting recursive structural types (and thus fully supporting structural
typing in these so-called hybrid languages) leads to undesirable consequences.
The interested reader is again advised to see~\cite{AbdelGawad2015}
for more details.}

\subsection{\label{sub:-Compared-toCOOK}$\protect\NOOP$ Compared to $\mu\protect\SOOP$}

Cook built on Cardelli's work %
by first developing a model of untyped inheritance~\cite{CookDenotational89,Cook1989}%
{} and, with others, then built a model of typed inheritance~\cite{CookInheritance90}.
In his work, Cook took self-referential classes, and thus recursive
types, in consideration, but, following the footsteps of Cardelli,
Cook kept a structural view of OO typing. Thus Cook et al. concluded
that `inheritance is not subtyping'~\cite{CookInheritance90}.%

Building on the work of Cook et al. and based on its conclusions,
Bruce, in his book on the foundations of OO languages~\cite{BruceFoundations02},
and Simons, in a series of articles on the theory of classification~\cite{SimonsTheory02},
enforced in the PL research community the conclusion reached by Cook
and his colleagues regarding breaking the relation between inheritance
and subtyping (implying the superiority of a structural view of OOP
in the process), even when the conclusion opposed and contradicted
the intuition (and even the ``conventional wisdom''~\cite[p.125]{CookInheritance90})
of a large section of OO developers and OO language designers. To
explain the discrepancy, it was then thought that mainstream OO languages
are technically deficient or flawed because, according to Cook~\cite{CookInheritance90},
these languages `place restrictions on inheritance'.

Given that $\mu\SOOP$\emph{ }(\emph{i.e.}, Cook et al's work) is
based on that of Cardelli, the differences between $\NOOP$ and $\SOOP$
we discussed in Section~\ref{sub:NOOPvsCardelli} get inherited by
a comparison between $\NOOP$ and $\mu\SOOP$.

The main technical similarity between $\NOOP$ and $\mu\SOOP$ is
that both models of OOP take self-referential classes, and thus recursive
types, in consideration. This is also where the two models strongly
disagree, since $\NOOP$ leads to a different conclusion about the
relation between inheritance and subtyping than $\mu\SOOP$ does.
This different conclusion is due to the differences in the nominal
view of objects versus the structural view of th em and to the inclusion/exclusion
of contracts in object typing and object subtyping, and accordingly
due to the role of inheritance (and thus contracts) in deciding subtyping.

As such, in addition to the main difference with $\SOOP$, comparing
$\NOOP$ to $\mu\SOOP$ highlights the following four differences,
which we first mention then discuss afterwards in some detail.
\begin{enumerate}
\item $\NOOP$ and $\mu\SOOP$ have different views of type names.
\item $\NOOP$ and $\mu\SOOP$ have different definitions of type inheritance.
\item $\NOOP$ and $\mu\SOOP$ are different as to the uniformity of their
inheritance models at the object level and at the type level.
\item $\NOOP$ and $\mu\SOOP$ are different as to the simplicity of the
mental model they present to developers during the OO software design
process.
\end{enumerate}
\noindent \emph{Views of type names} As we discussed, in detail, in
Section~\ref{sub:Type-Names,Circularity,Bin-Methods}, a main difference
between nominal typing and structural typing that is illustrated by
comparing $\NOOP$ to $\mu\SOOP$ is how type names are viewed in
nominal versus structural OO type systems, them having fixed meanings
in the first, while allowing their meanings to get rebound (upon inheritance)
in the latter.\smallskip{}

\noindent \emph{Definitions of inheritance} It is worthy to note
that the different conclusion reached by $\NOOP$ than that by $\mu\SOOP$
on the relation between inheritance and subtyping is based, in particular,
on how the two models differently define inheritance. Cook defines
inheritance as `a mechanism for the definition of new program units
by modifying existing ones in the presence of self-reference'~\cite{CookDenotational89}.
Cook also intentionally targets modeling the multiple levels of inheritance
that take place in OOP \emph{uniformly} (as we discuss below), having
a single model of inheritance that models type-level inheritance and
object-level inheritance. Applied to types, Cook's definition of
inheritance based on a structural view of types makes type inheritance
`a mechanism for the definition of new record type expressions by
modifying existing ones, in the presence of `self-type' '. On the
other hand, for the purpose of modeling nominally-typed mainstream
OOP with a nominal view of types (as in $\NOOP$), Cook's definition
of type inheritance has to be changed to `a mechanism for the definition
of new class signatures by adding member (\emph{i.e.}, field and method)
signatures to an explicitly-specified set of existing class signatures.'

In contrast to Cook's structural definition of type inheritance,
the nominal definition of type inheritance, first, disregards self-types
as having relevance in the definition, in agreement with the intuitions
of mainstream OO developers about the inheritance of class signatures
(where\emph{ }it is implied that nominal typing, with its fixed bindings
of type names, only presents an approximation to self-types). Secondly,
also in agreement with intuitions of mainstream OO developers, the
nominal definition of type inheritance stresses \emph{explicitness}
in specifying inheritance, making inheritance an intended relation
that is based on behavioral contracts and structure, not an accidental
relation based only on structure.\smallskip{}

\noindent \emph{Uniformity of inheritance models} Structurally-typed
OOP, as modeled by $\mu\SOOP$, uniformly applies the \emph{same}
model of inheritance (\emph{i.e.}, Cook's model~\cite{CookDenotational89})
at the level of values (\emph{i.e.}, objects) and at the level of
types. Using the same model at both levels requires rebinding the
self-variable, at the value level, and rebinding of the self-type-variable,
at the type level, upon inheritance.%
{} Nominally-typed OOP, and thereby $\NOOP$, on the other hand, uses
\emph{two} different models of inheritance, one at the level of values
(\emph{i.e.}, objects) and another at the level of types. The model
of inheritance at the level of values used in nominally-typed OOP
(the model of~\cite{CookDenotational89} applies well) allows for
rebinding the self-variable upon inheritance. At the level of types,
however, a different model where type names do not get rebound is
used by nominally-typed OOP, since %
there is no exact notion of a self-type-variable in nominally-typed
OO languages (but only an approximation to it, using a superclass
name, is available, as we explain in Section~\ref{sub:Type-Names,Circularity,Bin-Methods}).

As such, while the model of inheritance used in $\mu\SOOP$ uniformly
applies to object-level inheritance and type-level inheritance, we
can see that the models of inheritance used in $\NOOP$ reflect the
non-uniformity of inheritance models in mainstream nominally-typed
OOP, where a different model (and thus a different definition of inheritance)
is used at the object level than that at the type level.\smallskip{}

\noindent \emph{Economy of OO software design conceptual model} Agreeing
with the intuitions and conventional wisdom of mainstreaom OOP software
developers and OOP language designers, $\NOOP$ proves that `inheritance
is subtyping'~\cite{NOOP,NOOPsumm,InhSubtyNWPT13}, \emph{i.e.},
that there is a one-to-one correspondence between OO type inheritance
and OO subtyping, while $\mu\SOOP$ breaks the correspondence and
proves that `inheritance is not subtyping'~\cite{CookInheritance90,BruceFoundations02,SimonsTheory02}.
Splitting inheritance from subtyping, as $\mu\SOOP$ necessitates,
requires a structurally-typed OOP developer to keep\emph{ two }hierarchies
in mind when developing his software, namely, the inheritance hierarchy
and the subtyping hierarchy\footnote{Bruce, in an attempt to address this issue, suggested that OO languages
replace subtyping with `match-bounded polymorphism' (which is a
simplification of F-bounded polymorphism~\cite{CanningFbounded89,Baldan1999})
then identify type inheritance with matching. Matching~\cite{BruceBinary94},
upon which match-bounded polymorphism depends, however, uses subtyping
in its definition. As such, match-bounded polymorphism is not truly
a full replacement of subtyping, since developers still need to understand
subtyping to be able to understand matching. Having a non-simple mental
model of OOP, due to insisting on maintaining the split between subtyping
and inheritance, creates significant conceptual problems when designing
OO software. We speculate that this led Bruce's suggested language
extensions on matching to not gain traction or support in mainstream
OO languages.}.

This complexity, and the disregard of class contracts in deciding
subtyping, creates significant problems from the perspective of OO
program design (See~\cite{AbdelGawad2015}). Respecting semantic
class contracts in subtyping (thereby maintaining the identification
of inheritance with subtyping) allows nominally-typed OOP developers
on the other hand to reason about their software more readily and
to keep only \emph{one }hierarchy in mind while developing their software,
leading to a simpler more economic software design conceptual model.

Table~\vref{tab:NOOPvsSOOPvsmuSOOP} summarizes the similarities
and differences between $\NOOP$, $\SOOP$ and $\mu\SOOP$%
.

\noindent 
\begin{table}
\noindent \begin{centering}
\begin{tabular}{|c|c|c|c|}
\cline{2-4} 
\multicolumn{1}{c|}{} & $\SOOP$ & \noun{$\mu\SOOP$} & $\NOOP$\tabularnewline
\multicolumn{1}{c|}{} & (Cardelli; 1980s) & (Cook et al; 1990s) & (AbdelGawad; 2010s)\tabularnewline
\hline 
 & Structural model; & Structural model;  & Nominal model;\tabularnewline
\textsl{Nominal Info.} & Class names info. & Class names info. & Full class names info.\tabularnewline
 &  missing from objects & missing from objects &  included in objects\tabularnewline
\hline 
\multirow{2}{*}{\textsl{Object Types}} & Structural (reflect & Structural (reflect & Nominal (reflect struc.\tabularnewline
 & only object structure) & only object structure) & and assoc. contracts)\tabularnewline
\hline 
\textsl{Recursive Types} & Excluded & Included & Included\tabularnewline
\hline 
\textsl{View of} & Shortcuts. Self-ref. & Shortcuts. Self-ref. & Associated with public\tabularnewline
\textsl{Type Names} & not considered & gets rebound & contracts. No rebinding\tabularnewline
\hline 
\textsl{Inheritance} & Redefine inheritance & Same rebinding & Object level: Re-\tabularnewline
\textsl{Models} & as non-recursive & model at object & binding. Type\tabularnewline
 &  structural subtyping & level and type level & level: No rebinding\tabularnewline
\hline 
\textsl{Type Inheritance} & Structural & Structural & Nominal\tabularnewline
\hline 
\textsl{Conceptual} & \multirow{2}{*}{Inher. $=$ Subty.} & \multirow{2}{*}{Inher. $\neq$ Subty.} & \multirow{2}{*}{Inher. $=$ Subty.}\tabularnewline
\textsl{Economy} &  &  & \tabularnewline
\hline 
\end{tabular}
\par\end{centering}

\noindent \centering{}\protect\caption{\label{tab:NOOPvsSOOPvsmuSOOP}$\protect\SOOP$ vs. $\mu\protect\SOOP$
vs. $\protect\NOOP$}
\end{table}

\section{\label{sec:Conclusion}Concluding Remarks\label{sec:Future-Work}
and Future Work}

The identification of types with behavioral contracts, and of subtyping
with the inheritance and possible narrowing of contracts, makes nominal
typing and nominal subtyping in nominally-typed OOP closer to semantic
typing and semantic subtyping. Based on noting that%
, in this paper we compared a nominally-typed domain-theoretic model
of OOP to the most well-known structurally-typed models. Our comparison
has shown that nominally-typed models and structurally-typed models
of OOP lead to different views of fundamental notions of object-oriented
programming, namely objects, type names, class types, subtyping and
the relation between subtyping and inheritance.

In particular, our comparison highlights that in nominally-typed OOP
\begin{enumerate}
\item An object should not be mathematically viewed as merely a records
of its members (\emph{i.e.}, its fields and methods) but rather as
\textbf{a record together with nominal information} that is associated
with class contracts that the object maintains---this information
being carried along with the record, behaviorally constraining its
members,
\item A class type should not be viewed as a record type but rather as \textbf{a
record type that additionally respects behavioral contracts} associated
with nominal information embedded in elements of the type (\emph{i.e.},
its objects), and
\item Inheritance is correctly identified with nominal subtyping, \emph{i.e.},
that in pure nominally-typed OOP \textbf{inheritance is subtyping}.
\end{enumerate}
We believe the development of $\mathbf{NOOP}$, and the mathematical
comparison presented in this paper, are significant steps in providing
a full account of the relation between nominal and structural OO type
systems. 

Further, we hope that having a more accurate mathematical view of
nominally-typed OO software presents programming languages researchers
with better chances for progressing mainstream OO programming languages.
For example, generics (\cite{JLS14,CSharp2015,Odersky14})  add
to the expressiveness of type systems of nominally-typed OO programming
languages (\cite{Bank96,Agesen97,Bracha98,FJ/FGJ}). As hinted to
earlier, we believe that F-bounded generics offer better support for
binary methods in nominally-typed OO languages while maintaining the
benefits of nominal typing. Building a domain-theoretic model of generic
nominally-typed OOP, akin to $\NOOP$, and comparing it to domain-theoretic
models of polymorphic structurally-typed OOP, can as such offer better
chances for having a deeper understanding of features of generic mainstream
OO languages such as generic binary methods, variance annotations
(such as Java wildcards), Java erasure, polymorphic methods, and generic
type inference.

\section*{Acknowledgments}

The author expresses his gratitude and appreciation to Professor Robert
``Corky'' Cartwright for the discussions we had and the guidance
he gave that helped in developing and reaching some of the conclusions
in this paper, and to Professor Benjamin Pierce for the feedback he
offered on motivating and presenting $\NOOP$%
.

\bibliographystyle{plain}

\end{document}